\documentclass[floatfix,prl,showpacs,twocolumn]{revtex4}
\usepackage{graphicx,amsmath,bbm,mathrsfs,amssymb,pstricks,times,psfrag}
\newcommand{\ket}[1]{\vert #1 \rangle} \newcommand{\bra}[1]{\langle #1 \vert}
\newcommand{\thint}{\Theta_{\rm int}}
\newcommand{\Nex}{N_{\rm ex}} \newcommand{\Nth}{n_{\rm th}}
\newcommand{\Nit}{N_{\rm it}} \newcommand{\Nx}{N_{x}}
\begin{document}
\title{Reconstruction of the photon distribution in a micromaser}
\author{Stefano Olivares}
\email{stefano.olivares@mi.infn.it}
\author{Federico Casagrande}
\email{federico.casagrande@mi.infn.it}
\author{Alfredo Lulli}
\email{alfredo.lulli@unimi.it}
\author{Matteo G. A. Paris}
\email{matteo.paris@fisica.unimi.it}
\affiliation{Dipartimento di Fisica dell'Universit\`a di Milano, I-20133
Milano, Italia.}
\begin{abstract}
We suggest an iterative, maximum-likelihood-based, method to
reconstruct the photon number distribution of the steady state
cavity field of a micromaser starting from the statistics of the
atoms leaving the cavity after the interaction. The scheme is
based on measuring the atomic populations of probe atoms for
different interaction times and works effectively using a small
number of atoms and a limited sampling of the interaction times.
The method has been tested by numerically simulated experiments
showing that it may be reliably used in any micromaser regime
leading to high-fidelity reconstructions for single-peaked
distributions as well as for double-peaked ones and for trapping
states.
\end{abstract}
\date{\today}
\maketitle
\indent {\em Introduction} --- 
The one-atom maser or \textit{micromaser} is perhaps the most
relevant example of open quantum system in cavity quantum
electrodynamics (CQED) \cite{OVER}. Since its first experimental
realization \cite{Meschede} this system has allowed to
investigate many fundamental aspects in quantum optics. The
micromaser dynamics results from the interplay of a coherent
interaction between a beam of two-level atoms and a resonant
cavity mode in the microwave domain, as described by the
Jaynes-Cummings (JC) model \cite{JC}, and the dissipative process
due to the contact of the cavity with the environment. At the
steady state the radiation field inside the high-Q cavity may
show highly non-classical features, as for example sub-Poissonian
photon statistics \cite{REMPE3} or quantum collapses and revivals
\cite{REMPE1}. In addition, states characterized by a truncated
photon number distribution the, the so-called \textit{trapping
states} (TS) of the cavity field \cite{TS}, may be generated.
These states show up only at very low temperature and may be
affected by collective atomic interactions \cite{TSCOLL}. Under
suitable pumping conditions, photon distribution at the steady
state may also show two coexisting maxima, that is the signature
of first order phase-transitions \cite{Filipowicz}. Operating the
micromaser under pulsed regime and trapping conditions the
generation of Fock states has been also reported \cite{FS}. A
micromaser was implemented also on a two photon transition
\cite{TWOPHOTON} and more recently a microlaser was operated in
the optical regime \cite{microlaser}.
\par
A crucial aspect of the micromaser is that, in order to preserve
high Q values of the cavity, the cavity field is not accessible
to direct measurements. As a consequence, any information on its
properties must be inferred from the atoms leaving the cavity
after the interaction. In fact, the relation between the atomic
statistics and the properties of the cavity field has been
theoretically and numerically investigated \cite{Herzog-Cresser}
also including the back-action due to the atomic measurements
\cite{BRIEGEL}. On the other hand, to the best of our knowledge,
no method has been suggested to reconstruct the whole photon
distribution by exploiting the complete information carried by
the atoms leaving the cavity. In earlier experiments
\cite{REMPE3,REMPE1} the atomic statistics was obtained by
counting the number of excited (ground) atoms in a time interval
longer than the cavity lifetime, and then this frequency was
compared with the theoretical expression (see below) for the
experimental set of parameters. As a matter of fact, the photon
distribution was not reconstructed from the measurements.  In
experiments leading to TS \cite{TS} the steady state photon
distribution is composed only by few terms allowing a simple fit
of experimental data, whereas in the experiments to generate Fock
states \cite{FS} the cavity field state $|n\rangle$ is prepared
by a pulse of $n$ pump atoms and only one probe atom is measured
to obtain the atomic inversion that ideally involves only one
Rabi frequency. In this case, the advantage to measure only one
probe atom is that of avoiding the cavity field state reduction
due to repeated atomic measurements.
\par
In this letter, we suggest a method to reconstruct the whole
steady state photon distribution of the cavity field starting
from measurements of the statistics of probe atoms. The basic
idea is that atoms leaving the cavity after different interaction
times are carrying the complete information about the cavity
field itself. Indeed, the method is based on measuring the atomic
statistics for different interaction times and then estimating
the photon distribution using maximum-likelihood reconstruction.
As we will see, the method is very effective in any operating
regime of the micromaser and allows reliable reconstructions for
single-peaked distributions as well as for multi-peaked ones and
for trapping states. Remarkably, the method works effectively
starting from the statistics of a small number of atoms and a
limited sampling of the interaction times.  As a consequence, the
atoms used to probe the cavity field are only slightly perturbing
the steady state, which itself depends on the interaction time of
the pump atoms, {\em i.e.} the method can be used {\em on-line}
with experiments. We also notice that at the steady state, the
cavity field density matrix is diagonal in the Fock number basis,
and thus the reconstruction of the photon distribution
corresponds to the full quantum state reconstruction. On the
other hand, the characteristics of the micromaser spectrum
\cite{SPECTRUM} are related to the decay of off-diagonal elements
of the cavity field density matrix in the transient regime.
\par
\begin{figure}[htb]
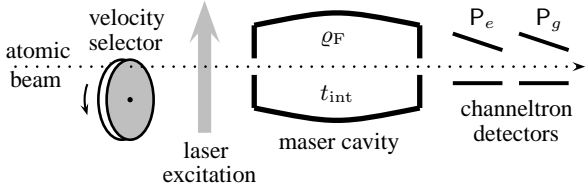

\begin{center}
\psset{unit=1.1cm} \pspicture(-1,0)(6,2)
\psbezier[linewidth=2pt]{-}(2,1.5)(3,1.7)(3,1.7)(4,1.5)
\psbezier[linewidth=2pt]{-}(2,0.5)(3,0.3)(3,0.3)(4,0.5)
\psline[linewidth=2pt](2,0.5)(2,0.9)
\psline[linewidth=2pt](4,0.5)(4,0.9)
\psline[linewidth=2pt](2,1.1)(2,1.5)
\psline[linewidth=2pt](4,1.1)(4,1.5)
\put(2.3,0){maser cavity}
\put(2.8,1.3){$\varrho_{\rm F}$}
\put(2.8,0.6){$t_{\rm int}$}
\put(-1,1.1){atomic}
\put(-0.95,0.75){beam}
\psline[linestyle=dotted,linewidth=1pt]{->}(-1,1)(0.4,1)
\psellipse[linewidth=1pt,fillcolor=white,fillstyle=solid](0.4,0.6)(0.3,0.5)
\qdisk(0.3,0.6){1pt}
\psellipse[linewidth=1pt,fillcolor=lightgray,fillstyle=solid](0.5,0.6)(0.3,0.5)
\qdisk(0.5,0.6){1pt}
\psarc{->}(0.5,0.6){0.55}{160}{200}
\put(0,1.5){velocity}
\put(0,1.2){selector}
\psline[linewidth=5pt,linecolor=lightgray]{->}(1.4,0.2)(1.4,1.8)
\put(1.15,-0.1){laser} \put(0.9,-0.4){excitation}
\psline[linestyle=dotted,linewidth=1pt]{->}(0.4,1)(6,1)
\psline[linewidth=1.5pt](4.4,0.8)(5,0.8)
\psline[linewidth=1.5pt](4.4,1.4)(5,1.2)
\psline[linewidth=1.5pt](5.2,0.8)(5.8,0.8)
\psline[linewidth=1.5pt](5.2,1.4)(5.8,1.2)
\put(4.6,1.5){${\sf P}_{e}$}
\put(5.4,1.5){${\sf P}_{g}$}
\put(4.5,0.4){channeltron}
\put(4.65,0.1){detectors}
\endpspicture
\end{center}
\caption{\label{f:probe} Schematic diagram of the micromaser setup.}
\end{figure}

 {\em Photon distribution at the steady state} ---
A schematic diagram of the micromaser setup is given in
Fig.~\ref{f:probe} where a beam of two level atoms, excited in the
upper Rydberg level of the maser transition, continuously and
resonantly pump a high-Q microwave cavity mode. The cavity
temperature is kept as low as $0.5 K$ in order to have a small
number of thermal photons. The velocity of the
atoms can be selected so that the interaction time $t_{int}$
between each atom and the cavity mode can be selected with high
precision. The atomic flux has a Poissonian distribution with a
mean pump rate $R$. The state of the atoms leaving the cavity can
be detected by field ionization techniques.
\par
The atomic decay rate $\gamma_a$ and the cavity decay rate
$\gamma$ are taken such that $t_{\rm int}\ll R^{-1}\ll
\gamma^{-1}\ll \gamma_a^{-1}$. Under the above conditions only one
atom interacts with the cavity mode each time, thus realizing a
perfect JC interaction, and a steady state regime for the cavity
field can be obtained. If $\varrho_{\rm F}$ denotes the (diagonal)
steady state density operator of the cavity field, the photon
distribution $p_n \equiv p_n(\thint,\Nex,\Nth)
=\bra{n}\varrho_{\rm F}\ket{n}$ can be expressed as
\cite{Filipowicz}
\begin{align}
p_n = p_0 \prod_{m=1}^{n} \frac{\left(\Nex / m\right)
\sin^2 \left( \thint \sqrt{m / \Nex} \right) + \Nth}
{1+\Nth}\:, \label{ph:stat}
\end{align}
where $p_0$ is a normalization constant, $\Nex=R/\gamma$ the
effective pump rate, and $\thint\equiv g t_{\rm int} \sqrt{\Nex}$
the dimensionless pump parameter, $g$ being the atom-cavity
coupling constant.
\par
A striking consequence of Eq.~(\ref{ph:stat}) is the existence of
trapping states of the cavity field \cite{TS}. In the limit of
$\Nth\rightarrow 0$ the distribution $p_n $ vanishes at photon
numbers $n_q$ ($q=1,2,...$) such that $ \thint \equiv q\pi
\sqrt{\Nex/(1+n_q)}$.  The TS correspond to narrow dips which
appear in the stationary mean photon number $\langle
N\rangle=\sum_n n p_n$ as a function of the pump parameter
$\thint$.  Another interesting form of $p_n$ can be obtained if
the pump parameter is set to $\thint\cong\frac{\pi}{2}$
corresponding to maximum amplification (MA) regime of the
micromaser. In this case, $p_n$ has a shape like that of a
coherent state with the same mean photon number. Finally, close to
$\thint=2\pi$ and multiples thereof, the photon distribution $p_n$
assumes a double-peaked (DP) structure corresponding to a
first-order phase transition \cite{OVER,Filipowicz}.
\par
When the system is at steady state, the probability ${\sf P}_e$
to find one atom in the excited state after its interaction with
the cavity field for a time $t_k$ is given by
\begin{equation}\label{th:Pe:bins}
{\sf P}_k = \sum_{n=0}^{\infty}
c_{kn}\,p_n\,, \quad  c_{kn} =\frac{1 + \cos\left(
\tau_k \sqrt{n+1} \right)}{2}\nonumber
\end{equation}
where ${\sf P}_k \equiv {\sf P}_e (\tau_k)$ and $\tau_k=gt_k$, 
is the dimensionless interaction time.  Eq.
(\ref{th:Pe:bins}) provides a link between the experimentally
measurable statistics of the probe atoms and the (inaccessible)
photon distribution of the cavity field.
\par {\it Reconstruction of the photon distribution} ---
At a first sight, Eq. (\ref{th:Pe:bins}) seems to provide a scarce
piece of information about the photon distribution $p_{n}$ of the
micromaser. However, if the atomic statistics is recorded for a
suitable set of values of the interaction times, then the
information is enough to reconstruct the full photon distribution.
As we will see, the inversion of Eq. (\ref{th:Pe:bins}), {\em
i.e.} the reconstruction of $p_n$, may be obtained by
maximum-likelihood estimation upon a suitable truncation of the
Hilbert space.
\par
The reconstruction scheme proceeds as follows: the micromaser is
pumped until it has reached the steady state for a fixed set of
parameters. Then we stop the atomic pump flux and a probe atom
prepared in the excited state is sent through the cavity in a time
much shorter than the cavity photon lifetime $\gamma^{-1}$.  The
velocity of this probe atom may be adjusted in order to vary the
interaction time in a given range. After the interaction with the
(steady state) cavity field the probe atom is detected.  We denote
by $f_k={\cal N}_k / N_x$ the experimental frequency of probe
atoms found in the excited state after an interaction time
$\tau_k$, $N_x$ being the total number of atoms sent through the
cavity with interaction time $\tau_k$. Of course, since atom
detection modifies the cavity field state, every probe atom is
followed by pump atoms to restore the steady state field.  In the
following, we assume that the values of interaction times for the
probe atoms $\tau_k$, $k=0,\ldots,n_\tau$ are uniformly
distributed between a minimum value $\tau_0$ and a maximum one
$\tau_{n_\tau}$, which, in turn, are determined by the maximum and
minimum velocities allowed by the specific experimental
implementation.
\par
Eq.~(\ref{th:Pe:bins}) is a statistical model for the parameters
$p_n$ that can be solved by maximum-likelihood (ML) estimation. We
assume that the photon distribution can be truncated at the
$\tilde{n}$-th term ({\em i.e.} $p_n$ is negligible for
$n>\tilde{n}$) and, without loss of generality, that $\Nx$ is
independent on $k$.  The global probability of the sample {\em i.e
} the log-likelihood (with normalized ${\sf P}_k$) of the detected
data reads as follows:
\begin{align}
{\sf L} = \frac{1}{\Nx} \log \prod_k
\left(\frac{{\sf P}_k}{\sum_m{\sf P}_m} \right)^{{\cal N}_k} \!\!\! =
\sum_k f_k \log \frac{{\sf P}_k}{\sum_m{\sf P}_m} \,.\label{log:lik}
\end{align}
ML estimates of $p_n$ are the values maximizing the log-likelihood
{\sf L}. Since the model is linear and the unknowns $p_n$ are
positive the solution can be obtained using an iterative procedure
\cite{kon,mogy,CVP,Hra}. Indeed, the equations $\frac{\partial
{\sf L}}{\partial p_n} = 0$ can be written as
\begin{equation}\label{quasi}
\frac{\sum_l {\sf P}_l}{\sum_l f_l}\, \sum_k \frac{c_{kn}}{\sum_m
c_{mn}} \frac{f_k}{{\sf P}_k} = 1\quad \forall
n=0,\cdots , \tilde{n}
\end{equation}
Then, by multiplying both the sides of Eq.~(\ref{quasi}) by
$p_n$, we get a map ${\boldsymbol T} p_n = p_n$, whose fixed
point can be obtained by the following iterative solution
\begin{equation}\label{EM:sol}
p_n^{(h+1)} = \frac{p_n^{(h)}}{\sum_m p_m^{(h)}}
\sum_k \frac{c_{kn}}{\left( \sum_l c_{ln} \right)} \,
\frac{f_k}{{\sf P}_e^{(h)}(\tau_k)}\,,
\end{equation}
where $p_n^{(h)}$ is the value of $p_n$ evaluated at the $h$-th
iteration, and ${\sf P}_e^{(h)}(\tau_k) = \sum_n
c_{kn}\,p_n^{(h)}$. Eq. (\ref{EM:sol}) is usually referred to as
the expectation-maximization solution of ML, and is known to
converge unbiasedly to the ML solution. As a matter of fact,
Eq.~(\ref{EM:sol}) provides a solution once the initial
distribution $p_n^{(0)}$ is chosen. In our simulated experiments
we start from the uniform distribution $p_n^{(0)}=(1+\tilde
n)^{-1}$ in the interval $[0,\tilde n]$, though any other
distribution $p_n^{(0)}$ such that $\sum_n p_n^{(0)}=1$,
$p_n^{(0)} \ne 0$ $\forall n$, would be appropriate as well.
Indeed, the initial distribution is slightly affecting only the
convergence rate and \emph{not} the precision at convergence
\cite{pcount}.
\begin{figure}[h!]
\pspicture(0,0)(8.5,13.0) 
\put(3.4,12.6){(a)} \put(0,8.8){
\includegraphics[width=0.22\textwidth]{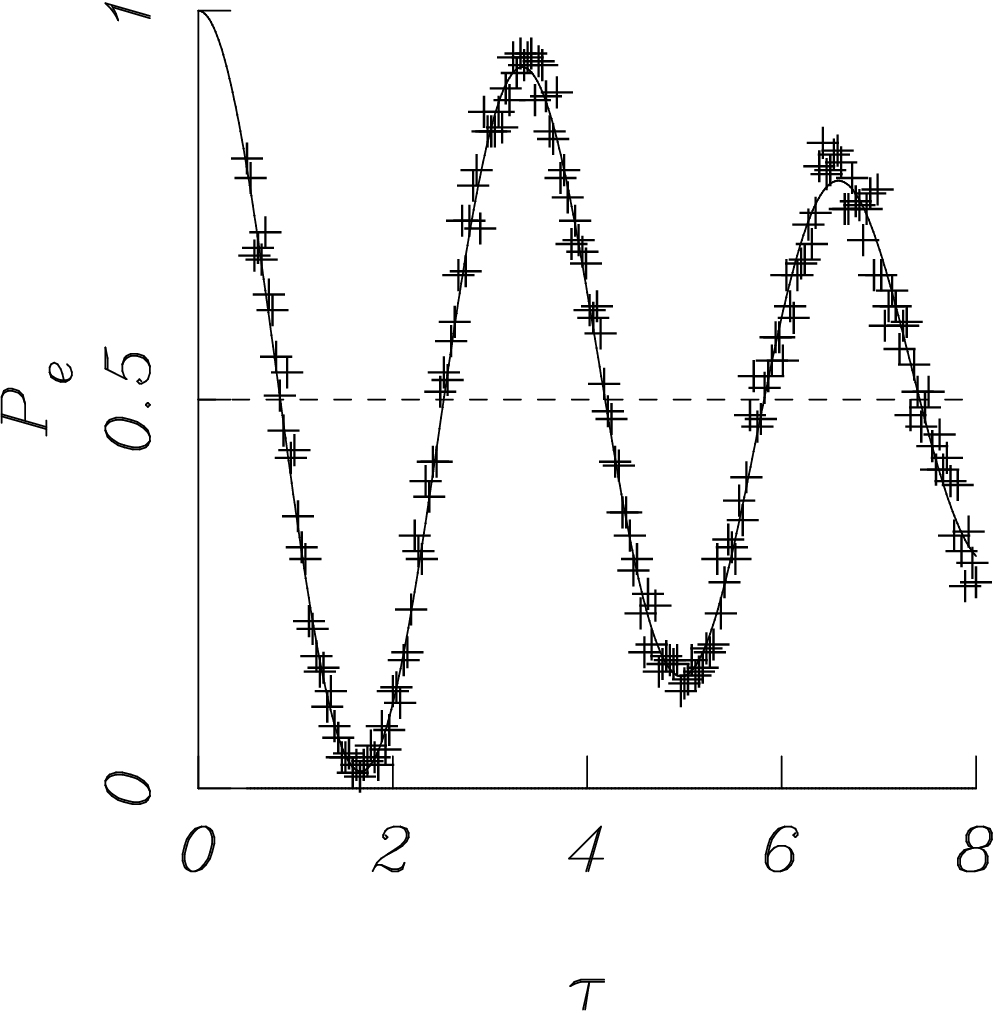} \hspace{0.2cm}
\includegraphics[width=0.22\textwidth]{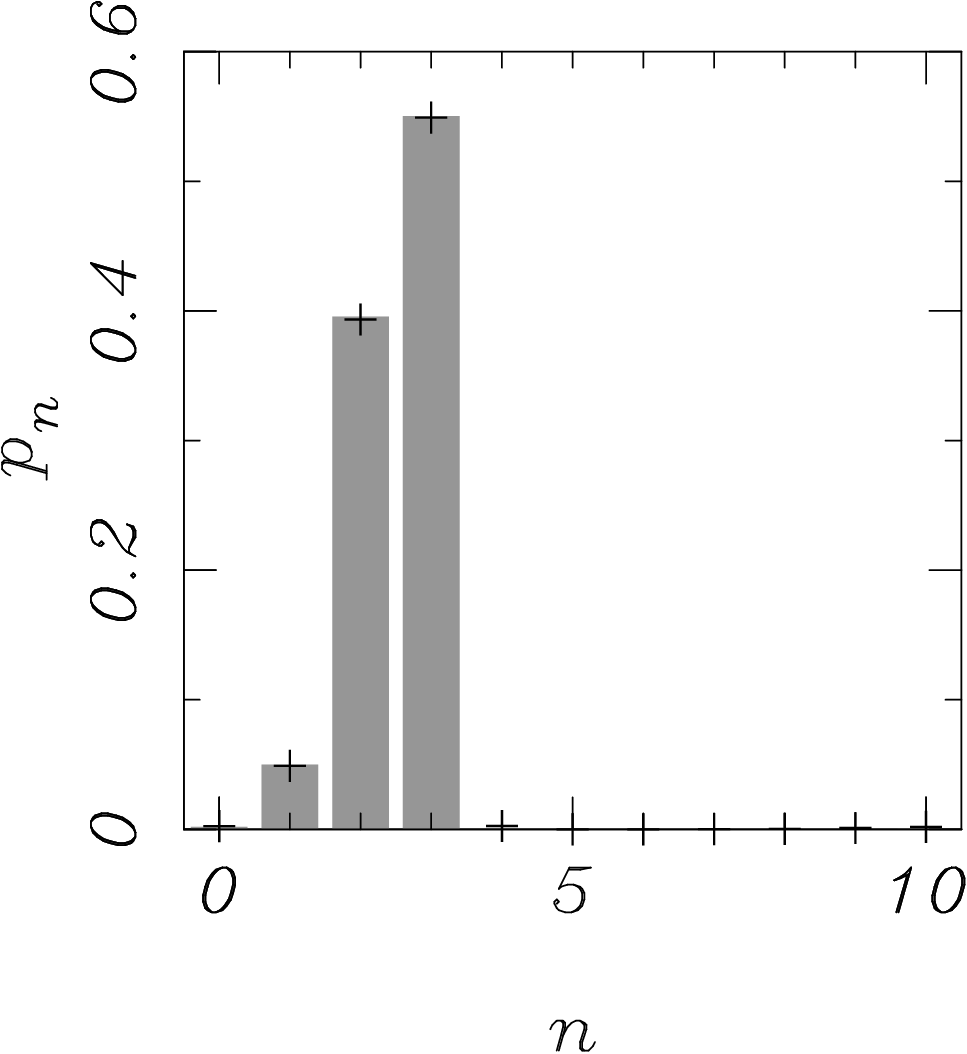}}
\put(3.4,8.2){(b)} \put(0,4.4){
\includegraphics[width=0.22\textwidth]{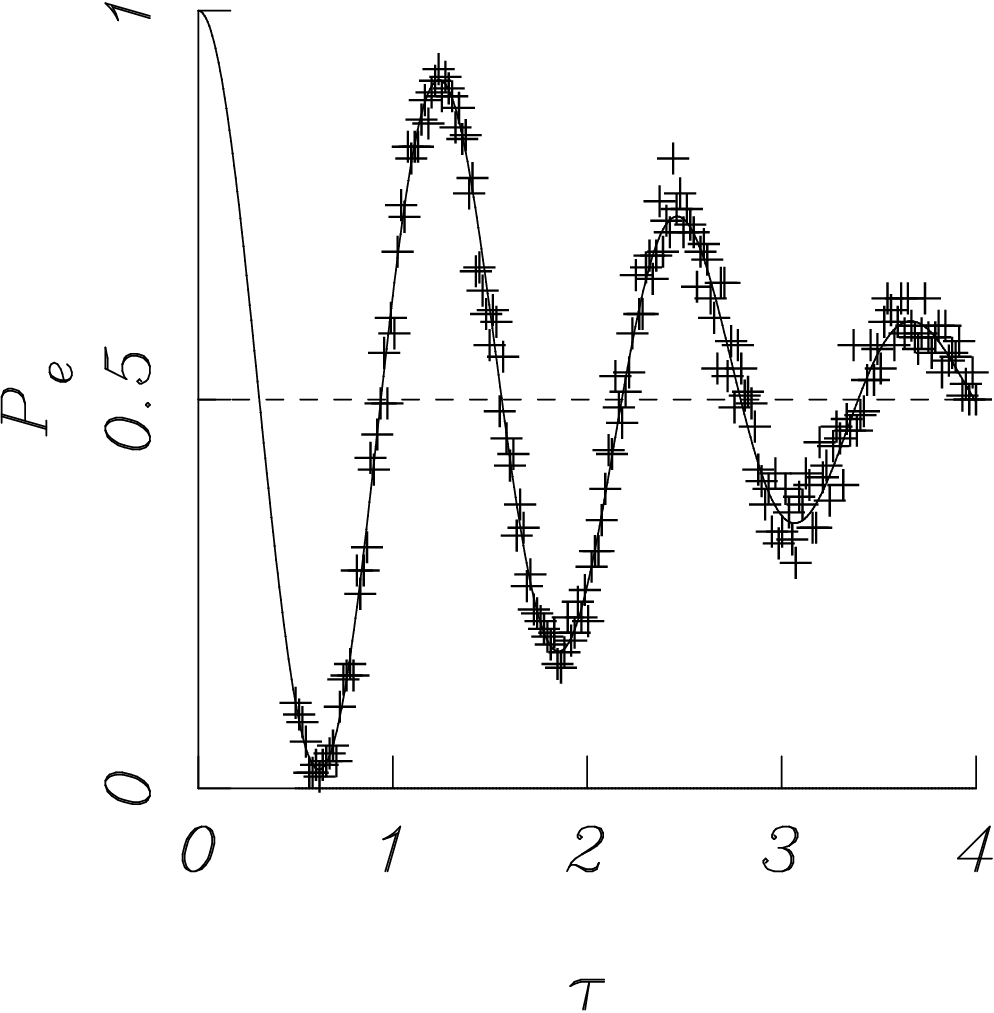} \hspace{0.2cm}
\includegraphics[width=0.22\textwidth]{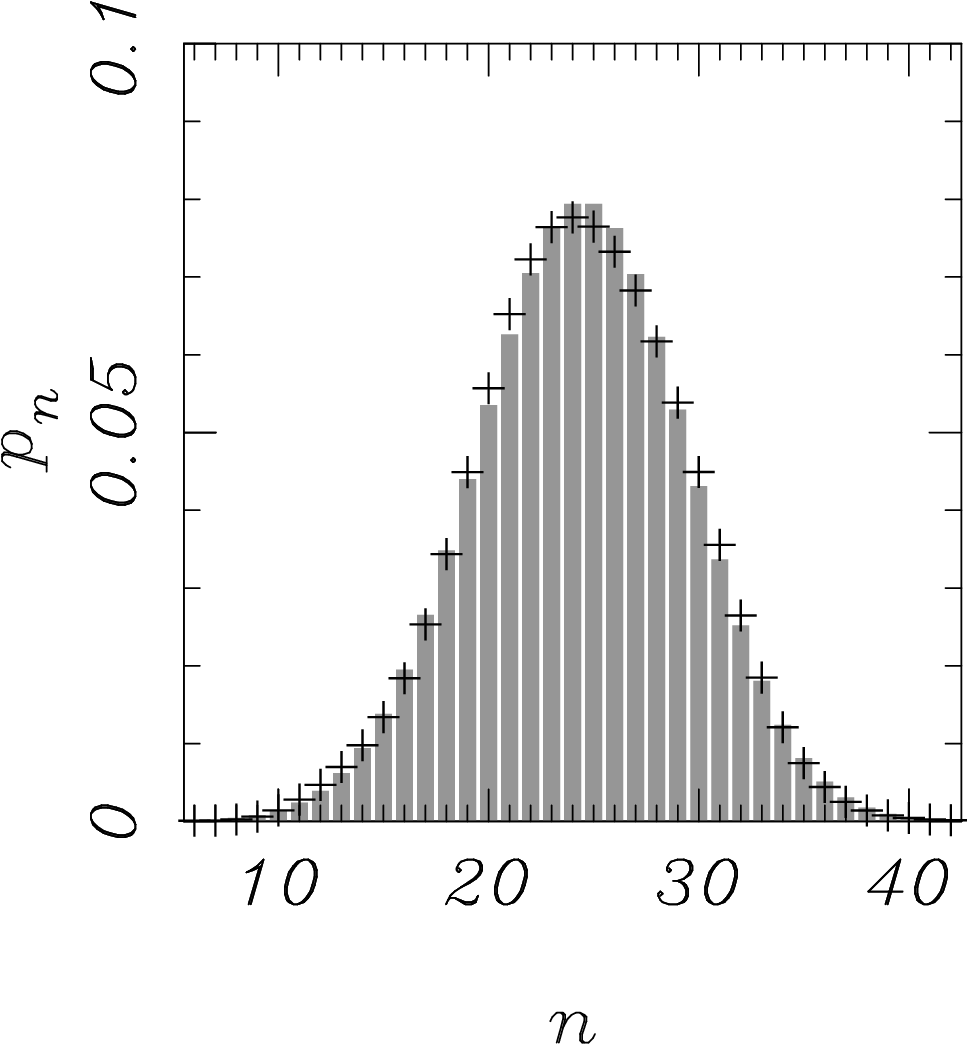}}
\put(3.4,3.8){(c)} \put(0,0){
\includegraphics[width=0.22\textwidth]{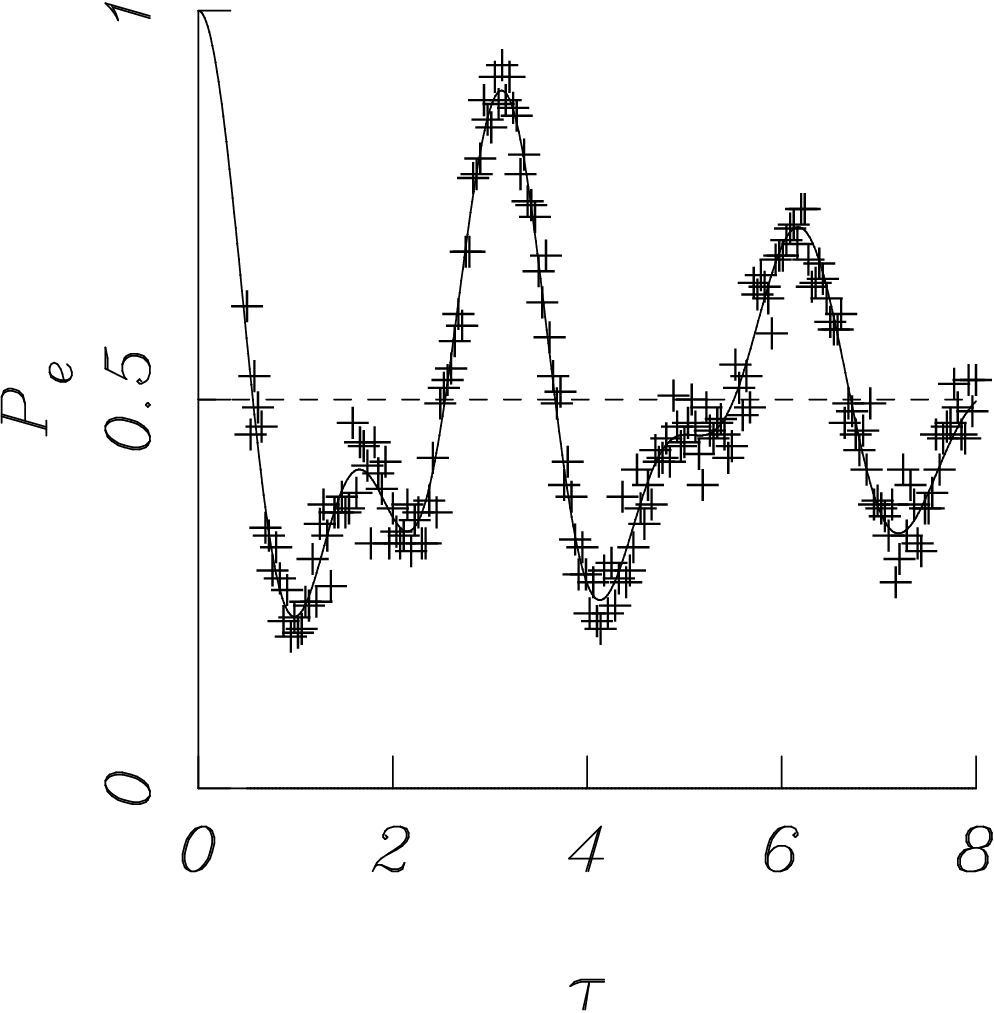} \hspace{0.2cm}
\includegraphics[width=0.22\textwidth]{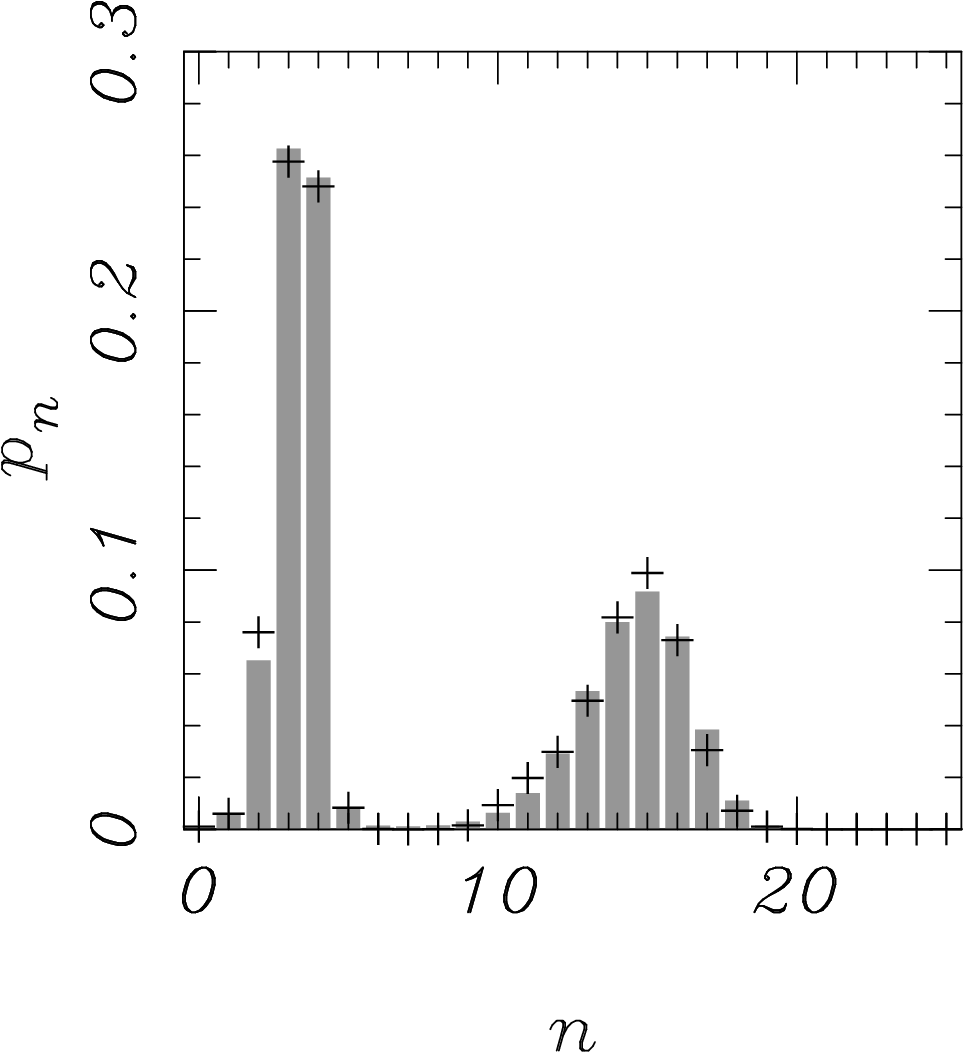}}
\endpspicture
\vspace{-0.3cm} \caption{\label{f:exp} Reconstruction of the
photon number distribution from Monte Carlo simulated experiments
for different steady state micromaser regimes. On the left we
report the probability ${\sf P}_e (\tau)$ of finding an atom in
the excited state as a function of the interaction time $\tau$ (as
obtained from the reconstructed distribution, solid line) compared
with the actual frequencies observed in the simulated experiments
(crosses).  On the right we show the reconstructed photon
distribution (crosses) compared with the theoretical one
(histograms).  The micromaser parameters are $\Nex=25.0$, $\Nth =
10^{-5}$ and (a) TS regime, $\thint/\pi = 2.5$; (b) MA regime,
$\thint/\pi = 0.5$; (c) DP regime, $\thint/\pi = 2.18$. In all the
simulated experiments $\tau_{0}=0.5$, $n_{\tau}=200$, and
$\Nx=200$, $\Nit=1000$.}
\end{figure}
\par {\em Monte Carlo simulated experiments} ---
Reliability and accuracy of the present method have been tested by
an extensive set of numerically simulated experiments,
corresponding to different micromaser steady state regimes. As a
figure of merit to assess the accuracy of the reconstructed
distribution $p_n^{\rm (r)}$, {\em i.e.} the similarity to the
actual distribution $p_n$ of Eq.~(\ref{ph:stat}), we consider the
fidelity $G = \sum_n \sqrt{ p_n^{\rm (r)}\,p_n}$. In
Fig.~\ref{f:exp} we show the simulated experimental data for the
measurement of ${\sf P}_e (\tau)$, generated by Monte Carlo
technique, and the comparison between the theoretical photon
distributions and those obtained by ML estimation. We consider as
interesting examples the TS, MA and DP steady state micromaser
regimes. In these regimes the photon number distribution is
sub-Poissonian, nearly Poissonian, and super-Poissonian,
respectively. In order to better appreciate the accuracy of our
reconstruction method we also report (see Table~\ref{t:exp}) the
first two moments of the cavity field distribution, {\em i.e.} the
mean photon number $\overline{n} = \langle a^\dag a\rangle$ and
the Fano factor $F = \left[\langle (a^\dag a)^2\rangle -\langle
a^\dag a\rangle^2\right]/\langle a^\dag a\rangle$, $a$ being the
mode operator of the cavity field and $\langle\cdots\rangle =
\hbox{Tr} \left[\varrho_{\rm F}\cdots\right]$ denoting ensemble
average.  As it is apparent from Table~\ref{t:exp} a very good
agreement is obtained for all the considered regimes between the
values obtained from the reconstructed distributions and the
actual ones.
\begin{table}[h!]
\caption{\label{t:exp} Mean photon number and Fano factor
of the reconstructed distributions of Fig.~\ref{f:exp} compared with the
theoretical values.}
\begin{ruledtabular}
\begin{tabular}{lllllll}
&$\thint$ & $\overline{n}$ & $F$ & $\overline{n}^{\rm (r)}$ &
$F^{\rm (r)}$ & $G$ $(\%)$ \\
TS &2.5$\,\pi$ & 2.52 & 0.22 & 2.53 & 0.21 & 99.73 \\
MA &0.5$\,\pi$ & 24.38 & 1.02 & 24.34 & 1.04 & 99.94 \\
DP &2.18$\,\pi$ & 7.85 & 4.05 & 7.76 & 4.03 & 99.72
\end{tabular}
\end{ruledtabular}
\end{table}
\par
\begin{figure}[h]
\pspicture(0,0)(8.5,4) \put(1,3.3){(a)}
\put(5.3,3.3){(b)} \put(0,0){
\includegraphics[width=0.22\textwidth]{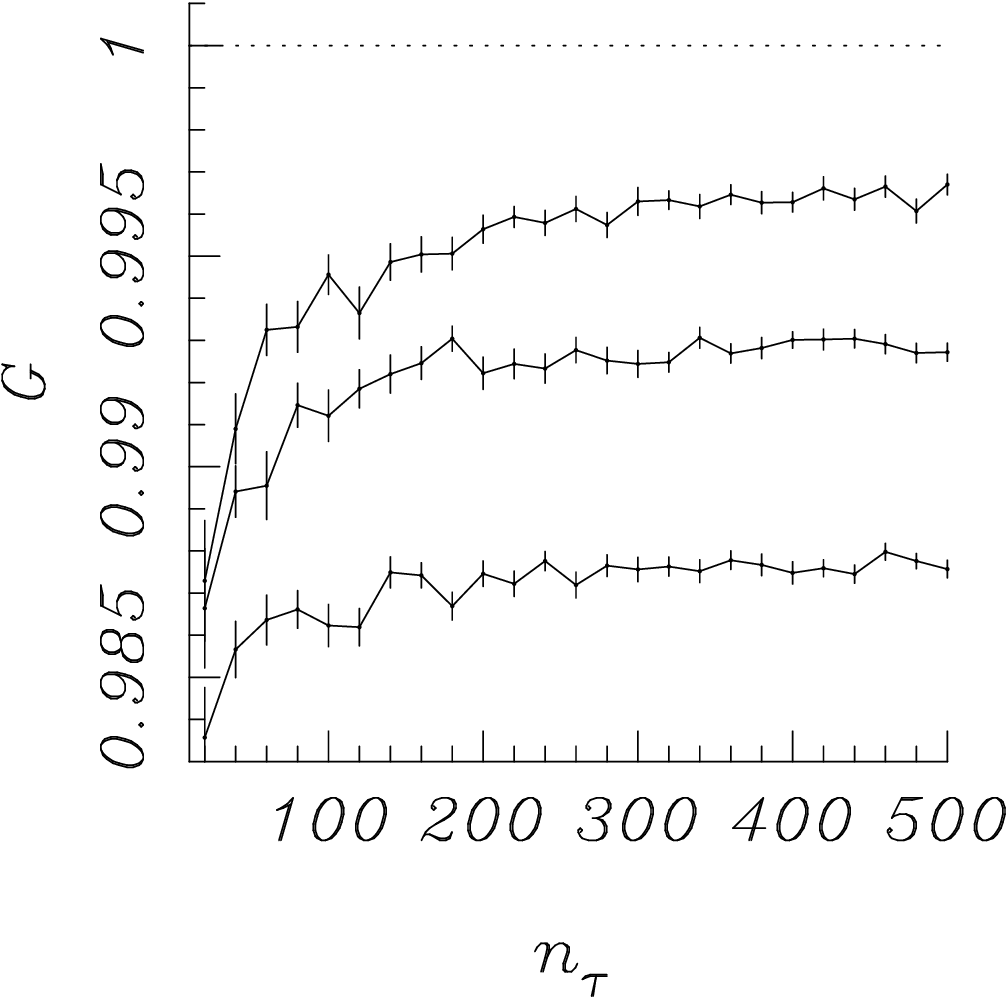} \hspace{0.2cm}
\includegraphics[width=0.22\textwidth]{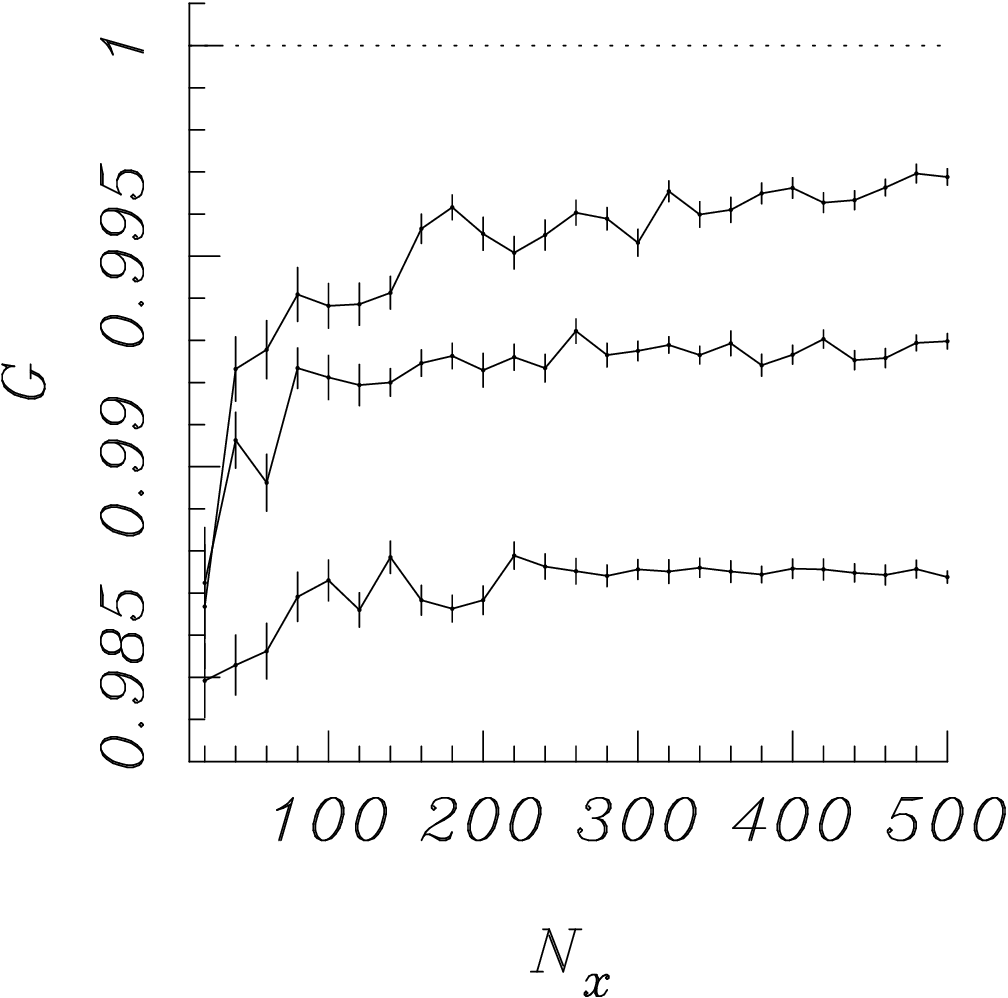} }
\endpspicture
\vspace{-0.3cm}
\caption{\label{f:conv:EM} (a): Fidelity of the reconstruction $G$ as
a function of the number $n_\tau$ of sampling times at fixed
number of data $\Nx=200$ for each time value. (b) Fidelity of the
reconstruction $G$ as a function of the number $\Nx$ of data for
each time value at fixed number $n_\tau=200$ of sampling
interaction times. Both plots refer to the case of TS state, {\em
i.e.} the reconstruction reported in Fig.~\ref{f:exp}a.  In both
plots we report the fidelity for different values of the number of
iterations $\Nit$, from bottom to top: $\Nit=100,200$ and $1000$.
The error bars are obtained by averaging over one hundred
simulated experiments.}
\end{figure}

Being our reconstruction method based on an iterative solution an
important aspect to keep under control is its convergence.  In
Fig.~\ref{f:conv:EM} we show the fidelity of the reconstruction
as a function of the number $n_{\tau}$ of sampling interaction
times  and the number $N_x$ of measures for each interaction
time, for different numbers of iterations $\Nit$.
As it is apparent from Fig.~\ref{f:conv:EM} the fidelity increases
with both $n_{\tau}$ and $\Nx$ and it reaches an asymptotic value
which actually depends on the choice of the other parameters. Of
course, also the number of iterations $\Nit$ affects the fidelity
value at convergence. Notice, however, that the reconstruction is
already very accurate with a number of iterations $\Nit=100$.  It
is worth noticing that the number of sampling times $n_{\tau}$
cannot be increased at will, since it is limited by experimental
constraints. In order to check the statistical reliability of the
algorithm we report the results from repeated (simulated) 
experiments. The error bars in Fig. \ref{f:conv:EM} are obtained 
by averaging over one hundred simulated experiments.
\par
\begin{figure}[h]
\includegraphics[width=0.22\textwidth]{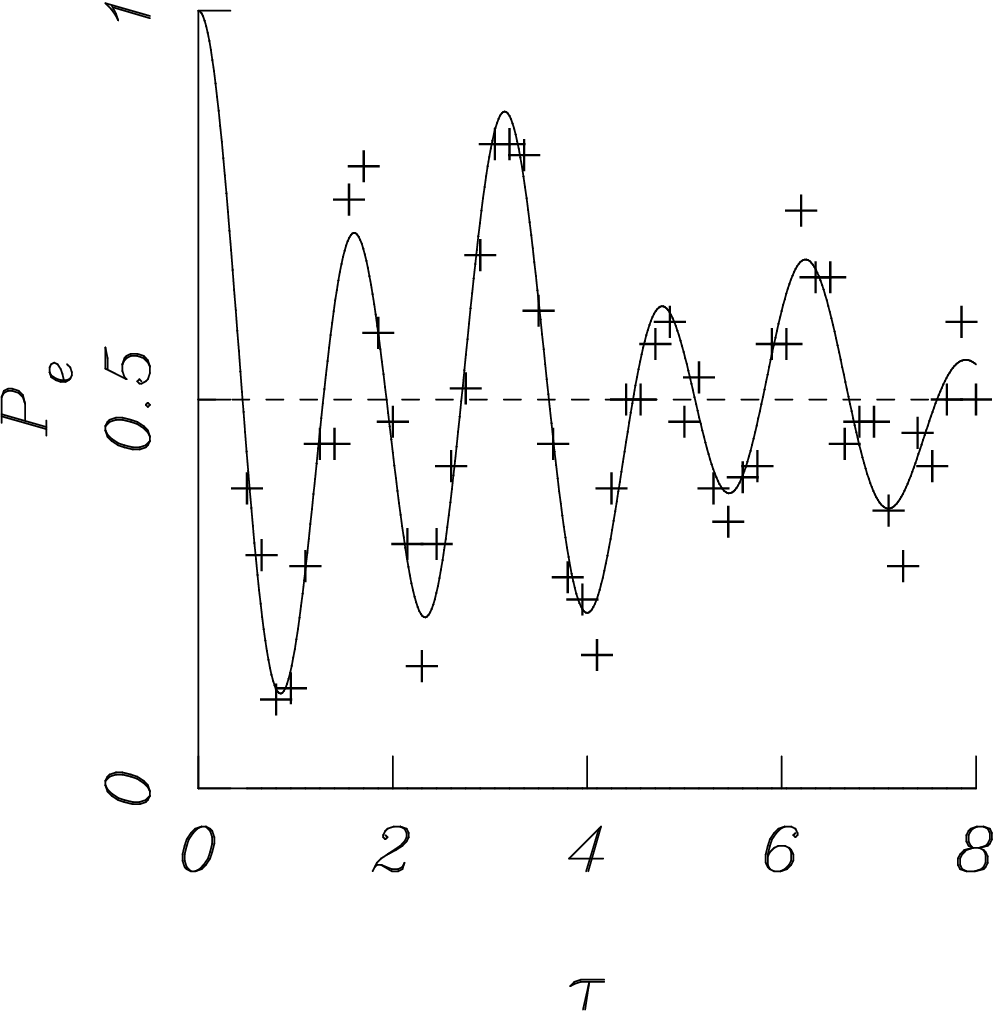} \hspace{0.2cm}
\includegraphics[width=0.22\textwidth]{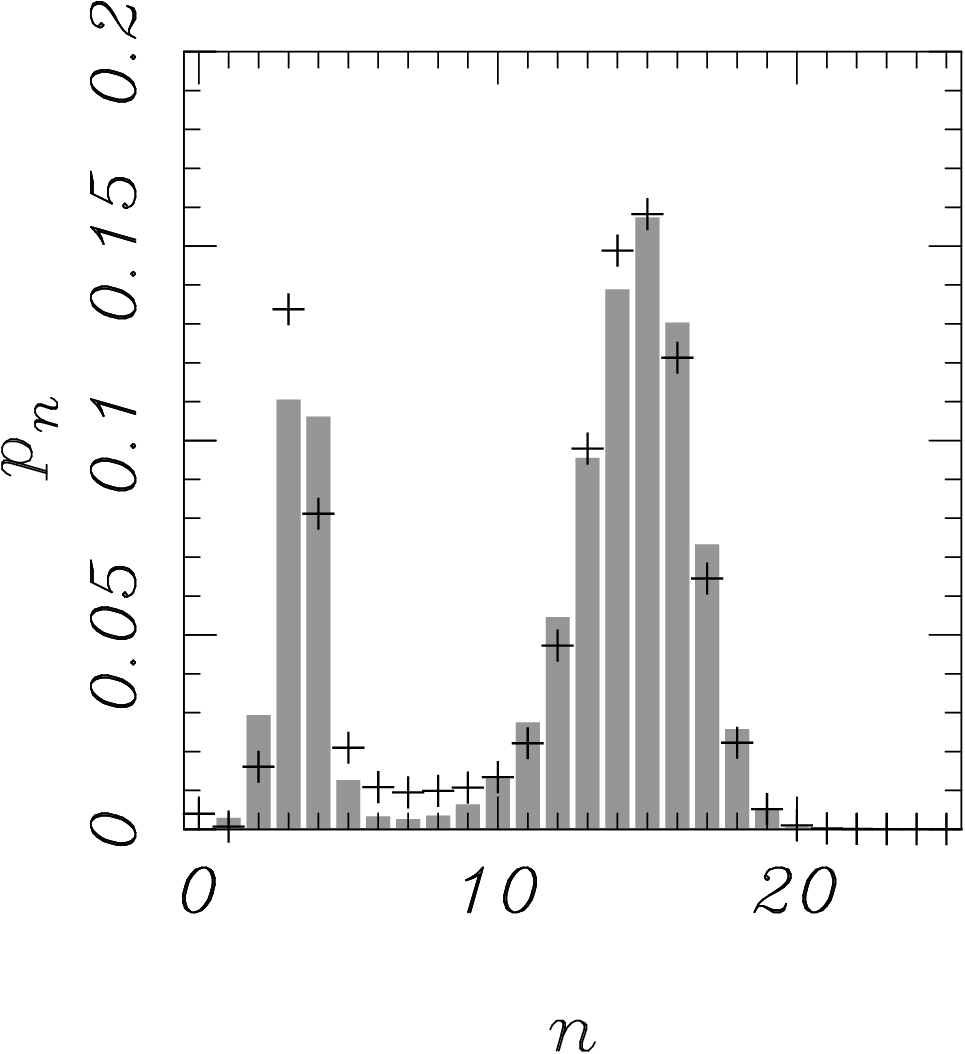}
\vspace{-0.3cm} \caption{\label{f:exp:bis} Reconstruction of the
photon number distribution as in Fig.~\ref{f:exp}(c) but with
reduced sampling parameters $N_x=70$, 
$n_{\tau}=50$, $\Nit=300$, and $\tilde
n = 25$. For these values of parameters we have $\overline{n}=
11.55$, $F= 2.34$, $\overline{n}^{\rm (r)} = 11.41$, $F^{\rm (r)}
= 2.35$, and $G = 98.14\%$.}
\end{figure}

A question arises on whether the present method
could be effectively employed with a small number of atoms and a
limited sampling of the interaction times. This is of course a
crucial aspect concerning its possible implementation in a
realistic scenario. We found, by means of an extensive set of
simulated experiments, that the answer is positive and that
accurate reconstructions may be obtained using realistic values of
the parameters.  In Figs.~\ref{f:exp:bis} we report, as an
example, the results of a simulated experiment, corresponding to
that of Fig.~\ref{f:exp}(c), now performed with $n_{\tau}=40$,
$\Nx=30$ and $\Nit=50$. As it is apparent from the plots, the
reconstruction is still very accurate despite the fact that the
total number of observations has been dramatically decreased.
\par{\em Summary and conclusions} ---
We have suggested a novel iterative method to reconstruct the full
photon distribution of the cavity field of a micromaser at the
steady-state starting from the statistics of the probe atoms
leaving the cavity after different interaction times. Our methods
works effectively using a small number of atoms and a limited
sampling of the interaction times.  This features, together with
its accuracy and fast convergence, make it suitable for being used
{\em on-line} with experiments. The method has been tested by
numerically simulated experiments showing that it may be reliably
used in any steady state regime of the micromaser leading to
high-fidelity reconstructions for single-peaked distributions as
well as for double-peaked ones and for trapping states.
\par\noindent\\
This work is dedicated to the memory of Herbert Walther. It has
been supported by MIUR project PRIN2005024254-002. MGAP is also
with ISI Foundation, Torino, Italy.

\end{document}